\documentclass[letterpaper, 10 pt, conference]{ieeeconf}  

\IEEEoverridecommandlockouts                              

\usepackage{epsfig} 

\title{\LARGE \bf
Upper extremity kinematics: Development of a quantitative measure of impairment severity and dissimilarity after stroke
}

\author{Khadija F. Zaidi$^{1}$ and Michelle Harris-Love$^{2}$
\thanks{*This work was supported by a Dissertation Completion Grant awarded by George Mason University}
\thanks{$^{1}$Khadija F. Zaidi is a Doctoral Candidate in the Computational Biomedical Engineering concentration of the Department of Bioengineering
        Volgenau School of Engineering, George Mason University, Fairfax, Virginia, USA
        {\tt\small szaidi8@gmu.edu}}%
\thanks{$^{2}$Michelle Harris-Love is with Anschutz Medical Campus, University of Colorado,
        Aurora, Colorado, USA
        }%
}

\begin{document}

\maketitle
\thispagestyle{empty}
\pagestyle{empty}

\begin{abstract}
\newline
\textbf{\textit{Background}}
Strokes are a leading cause of disability worldwide, with many survivors experiencing difficulty in recovering upper extremity movement, particularly hand function and grasping ability. There is currently no objective measure of movement quality, and without it, rehabilitative interventions remain at best informed estimations of the underlying neural structures' response to produce movement. In this paper, we utilize a novel modification to Procrustean distance to quantify curve dissimilarity and propose the Reach Severity and Dissimilarity Index (RSDI) as an objective measure of motor deficits.\newline
\textbf{\textit{Methods}} 
All experiments took place at the Medstar National Rehabilitation Hospital; persons with stroke were recruited from the hospital patient population. Using Fugl-Meyer (FM) scores and reach capacities, stroke survivors were placed in either mild or severe impairment groups. Individual completed sets of reach-to-target tasks to extrapolate kinematic metrics describing motor performance. The Procrustes method of statistical shape analysis was modified to identify reaching sub-movements that were congruous to able-bodied sub-movements.\newline
\textbf{\textit{Findings}} 
Movement initiation proceeds comparably to the reference curve in both two- and three-dimensional representations of mild impairment movement. There were significant effects of the location of congruent segments between subject and reference curves, mean velocities, peak roll angle, and target error. These metrics were used to calculate a preliminary RSDI score with severity and dissimilarity sub-scores, and subjects were reclassified in terms of rehabilitation goals as "Speed Emphasis", "Strength Emphasis", and "Combined Emphasis".. \newline
\textbf{\textit{Interpretation}}
The Modified Procrustes method shows promise in identifying disruptions in movement and monitoring recovery without adding to patient or clinician burden. The proposed RSDI score, while limited in scope, can be adapted and expanded to other functional movements and used as an objective clinical tool. By reducing the impact of stroke on disability, there is a significant potential to improve quality of life through individualized rehabilitation.
\end{abstract}

\flushbottom
\maketitle
\thispagestyle{empty}

\section*{Introduction}

Strokes represent one of the leading causes of disability worldwide. 65\% of stroke survivors experience some difficulty in recovering the ability to reach \cite{Krauth19, Mesquita19_Part2, Mesquita19_Part1}, with more severe impairments featuring a loss of hand function and ability to grasp \cite{Guzman14,Reissner19, Szwedo21}.At 6 months post stroke, many continue to experience some degree of upper extremity hemiparesis. This unilateral impairment of the paretic limb impacts functional reaching, and is a major contributor to stroke-related disability \cite{Mohapatra16}. 

Early signs of motor control interruption include paralysis, reduced reflexes, and inability to produce resistance to perturbations \cite{McCrea05, Sing13}. Symptoms arising during the chronic post-stroke recovery phase may include increased reflex activity or spasticity. \cite{Elliott20,HarrisLove12}. Compensatory movements may also arise in lieu of true recovery, such as extending the trunk to reach a target at arm's length due to decreased joint range of motion \cite{Saes22}. Stroke severity can significantly impact the type and amount of deficits experienced by an individual and the efficacy of particular rehabilitative strategies \cite{Cirstea00}. While mechanisms of arm recovery have been studied after mild functional impairments \cite{Morel17,Schwarz21}, there are few effective treatments for the large portion of the stroke population with more severe impairments. An objective measure of severity and the nature of deficits is of interest in creating individualized rehabilitation plans \cite{Wolff23,Yang18}. 

Three-dimensional kinematic analyses provide objective methods to characterize movement subsequent to stroke. \cite{Cai19, JarqueBou20, Schwarz19, Wolff22}. Kinematics of the upper extremity obtained through motion capture and 3D positional data can provide more sensitive tools to objectively assess individual motor function after stroke \cite{Scano19,Ozturk16, Jaspers11}. Active and passive visual markers, electromagnetic sensors, and inertial sensors have been used extensively for human movement analysis and can provide metrics such as movement speed, movement smoothness, joint angles, and limb orientation from position data \cite{Collins18, Ueyama21}. 


Currently, there is no consensus on the most appropriate tasks or variables to provide a global description of upper extremity movement \cite{Cacioppo20, Corona18, Andel08}. With significant variability between individuals, clinicians use measures such as the Upper Extremity Fugel-Meyer scale to subjectively describe movement capability \cite{Singer17, Woytowicz17}. Without an objective measure of movement quality, rehabilitative interventions are at best informed estimations of how the underlying neural structures will respond and produce movement. Subjective clinical scores cannot identify wherein during movement a deficit occurs and what that might suggest as the best rehabilitative plan \cite{Roberts19, Roberts20, Priot20}. Subjective scales also cannot efficiently monitor changes in impairment severity and dissimilarity over time. 

In this paper, we propose a modified Procrustes analysis method applied to groups of persons with stroke, differentiated by movement severity. Utilizing upper extremity endpoint data from these two groups, this method was used to identify movement behaviors and metrics that differentiate the mild and severe impairment groups. Finally, this study includes a preliminary severity and dissimilarity score of upper extremity movement that draws inspiration from scores such as the Gait Profile Score (GPS)\cite{Baker09}, or Gait Deviation Index (GDI) \cite{Schwartz08}. The GPS evaluates overall gait pathology severity based solely on kinematic data for a given individual, while the GDI identifies how much an individual's gait features deviate from a reference set of able-bodied data. A single measure of the overall quality of a upper extremity movement, overall severity, and dissimilarity from reference data would be of interest in informing clinical decisions.

The paper is organized as follows: 

\begin{itemize}
\item Validity of utilizing Procrustean Distance in Upper Extremity Analysis,
\item Objectives and hypotheses of applying a modified Procrustes analysis to endpoint data,
\item This study's inclusion and exclusion criteria for persons with stroke,
\item Clinical measures used to classify patients into mild and severe impairment groups,
\item Description of the experimental protocol and study methodology,
\item Definitions of kinematic metrics included in the data analysis,
\item Statistical tests performed to identify significant differences between subject groups,
\item Resulting quantitative measures of severity and dissimilarity that inform the proposed \\"Reach Severity and Dissimilarity Index" (RSDI)
\end{itemize}

\section*{Background}

In mathematics, the Euclidean distance between two points is the length of a line drawn between them. Root-Mean-Square Error (RMSE) is another method of quantifying how much one set of data differs from a reference set. Both Euclidean distance and RMSE have been used to construct measures of movement quality in the lower limb \cite{Baker09, Karamanidis16} and the upper limb \cite{Cerveri07, Dehbandi16, Riad11}. Additionally, Principle Component Analyis (PCA) is commonly used to simplify the interdependent data that is necessary to represent participating limb segments and joints, task requirements, and environmental constraints that produce any particular movement \cite{Guzik22}. Clinical decisions can then be based on an interpretation of the complex data. The validity of scores generated by quantifying differences between mean reference data and paretic movement data has been established in the field of rehabilitation \cite{Jaspers11APS,Hill22}. 

Procrustes Analysis is another such psychometric method of quantifying difference or dissimilarity between two sets of data \cite{kendall89}. Procrustes distance has recently garnered attention as a metric in both gait \cite{rida19,Kamal18,Anwary19} and upper extremity studies \cite{Passos23,saenen22,wong19,passos17}. Procrustes Analysis quantifies similarity of shape between two matrix sets and provides the linear transformation that would allow one curve to best conform to the other. More specifically, the Procrustes method compares each ith element of the subject curve to the ith element of the reference curve. This method generates a scaling factor b, an orthogonal rotation and reflection matrix T, and a translation matrix C, and a Procrustes distance d. Computing the Procrustes distance presents an interesting advantage in quantifying subject performance. Additionally, the scaling factor b can indicate a prolonged or truncated movement, while the ability to compare a reflected curve can allow comparison of right and left limb movements to the same reference curve \cite{seber09,bookstein97}. In addition to discrete kinematic landmarks, the variability across an entire movement can be assessed in order to extrapolate a subjective and sensitive representation of upper limb movement. 

In order to support the proposed RSDI score, we quantitatively identified segments of the forward reaching movement that showed the least deviation when compared to a reference curve representing stereotypical able-bodied reaching behavior. These segments of movement were characterized not by the magnitude of discrete kinematic metrics but rather by when they occur relative to those metrics and when during the overall movement. We hypothesize that subjects with mild impairment will exhibit initial acceleration behaviors that are analogous to healthy movement, while subjects with more severe impairment will not exhibit any congruous segments of movement and therefore result in higher severity and deviation scores. Further, it is expected subjects with severe impairment will demonstrate diminished ability to refine movement through less variability in endpoint orientation. This study suggests the specific sub-movements, in cases of mild and severe impairment, that remain congruous to healthy movement can allow quantification of impairment severity and inform targets for rehabilitation. 

\section*{Methods}

\subsection*{Participants}

Participants were recruited from the Medstar National Rehabilitation Hospital stroke patient population. Patients' stroke diagnoses were confirmed via Magnetic Resonance Imaging (MRI). All subjects completed written informed consent forms. This protocol was approved by the Medstar Rehabilitation Research Institutional Review Board under protocol number [947339-3]. 

Persons with stroke that were (1) at least eighteen years of age, (2) able to complete a reach-to-target task, (3) able to consent to the study and experienced no significant cognitive deficits (Mini-Mental State Examination score $>24$), and (4) six or more months post thromboembolic non hemorrhagic hemispheric or hemorrhagic hemispheric strokes were recruited for this study. 

Potential subjects were excluded if (1) they were less than 18 years of age, (2) stroke occurred less than 6 months before participation or affected both hemispheres, (3) stroke involved the cerebellum, brainstem, or did not spare primary motor and dorsal premotor cortices, (4) there was a history of craniotomy, neurological disorders (other than stroke), cardiovascular disease, or active cancer or renal disease (5) there was a history of orthopedic injury or disorder affecting shoulder or elbow function, or (6) they had had a seizure or taken anti-seizure medication in the past 2 years. 

\begin{table*}
\caption{Mild Impairment Stroke Survivor Demographics \\UEFM - Upper Extremity Fugl-Meyer;MMSE - Mini Mental State Examination}\label{tbl1}
\begin{tabular*}{1.0\linewidth}{@{\extracolsep{\fill}}p{0.8cm}p{1cm}p{0.8cm}p{1cm}p{1cm}p{1.3cm}p{1.1cm}p{0.8cm}p{0.8cm}}
\hline
Sub \# & M/F & Age & Months Since Stroke & Paretic Arm & Dominant Affected & Max Paretic Reach (cm) & UEFM & MMSE\\
\hline
1	& M & 62 &	107	&	R	&	Y	& 47.4 & 59 & 30\\
2	& M & 64 &	72	&	R	&	Y	& 46.5 & 51 & 30\\
3	& M & 44 &	14	&	L	&	N	& 43.0 & 46 & 30\\
4	& M & 64 &  14	&	L	&	N	& 36.3 & 43 & 27\\
5	& M & 54 &	13	&	R	&	Y	& 40.8 & 42 & 27\\
6	& M & 59 &	68	&	L	&	N	& 27.2 & 54 & 26\\
7	& F & 57 &	22	&	L	&	N	& 20.5 & 64 & 27\\
8	& M & 60 &	48	&	L	&	N	& 29.2 & 63 & 26\\
9	& M & 44 &	42	&	L	&	Y	& 41.5 & 64 & 30\\
10	& M & 73 &	8	&	R	&	Y	& 19.6 & 50 & 25\\
11	& M & 77 &	55	&	R	&	Y	& 44.8 & 43 & 28\\
12	& F & 74 &	7	&	R	&	Y	& 32.1 & 61 & 26\\
13	& M & 65 &	78	&	L	&	N	& 29.2 & 53 & 27\\
14	& F & 59 & 	20	&	R	&	Y	& 37.5 & 39 & 30\\
15	& F & 71 & 	17	&	L	&	N	& 37.8 & 54 & 27\\

$Mean \pm SD$ & (M/F) = (11/4) & $61.8 \pm 9.8$ & $39 \pm 31$ & & (Y/N) = (9/6)& & $52.4 \pm 8.5$ \\

\end{tabular*}
\end{table*}

\begin{table*}
\caption{Severe Impairment Stroke Survivor Demographics}\label{tbl2}
\begin{tabular*}{1.0\linewidth}{@{\extracolsep{\fill}}p{0.8cm}p{1cm}p{0.8cm}p{1cm}p{1cm}p{1.3cm}p{1.1cm}p{0.8cm}p{0.8cm}}
\hline
Sub \# & M/F & Age & Months Since Stroke & Paretic Arm & Dominant Affected & Max Paretic Reach (cm) & UEFM & MMSE\\
\hline
1	&	F	&	69	&	12	&	L	&	N	&	13.8 &	10	&	29	\\
2	&	M	&	57	&	120	&	R	&	N	&	30.2 &	24	&	27	\\
3	&	M	&	56	&	16	&	L	&	Y	&	14 &	8	&	30	\\
4	&	M	&	63	&	11	&	L	&	N	&	27.8 &	29	&	25	\\
5	&	F	&	68	&	112	&	L	&	N	&	43.4 &	23	&	30	\\
6	&	F	&	44	&	30	&	R	&	N	&	22 &	25	&	24	\\
7	&	F	&	69	&	9	&	L	&	N	&	27 &	14	&	30	\\
8	&	M	&	51	&	5	&	L	&	N	&	19.2 &	10	&	29	\\
9	&	M	&	54	&	43	&	R	&	N	&	4.1 &	7	&	26	\\
10	&	F	&	70	&	401	&	R	&	Y	&	4 &	14	&	25	\\
11	&	F	&	78	&	8	&	R	&	Y	&	5.3 &	12	&	28	\\
12	&	F	&	63	&	25	&	L	&	N	&	14 &	22	&	28	\\
13	&	M	&	71	&	28	&	L	&	N	&	7.3 &	13	&	29	\\
14	&	M	&	57	&	49	&	R	&	Y	&	4 &	16	&	28	\\

$Mean \pm SD$ & (M/F) = (7/7) & $62.1 \pm 9.3$ & $62.1 \pm 104.2$ & & (Y/N) = (4/10) & & $16.2 \pm 7.1$ \\

\end{tabular*}
\end{table*}


\subsection*{Clinical Measures}

Demographics for participants with mild and severe impairments after stroke are detailed in Tables \ref{tbl1} and \ref{tbl2}. 


Subjects underwent a Mini-Mental State Examination \cite{Folstein75} to ensure ability to consent to all sections of the study and complete tasks as instructed. Since this study features a functional reaching task for the upper extremity only, assessment of recovery was limited to the Upper Extremity Motor Function section of the Fugl-Meyer Assessment. The Upper Extremity Fugl-Meyer (UEFM) test was used as a criterion for classifying post-stroke impairment as either mild or severe upper limb impairment. Classifications for impairment severity have been proposed in prior literature based on a range of Motor Function scores, \cite{FMmethod, Duncan94}. The Motor Function domain is divided into the following: Upper Extremity (scored out of 36), Hand (scored out of 10), Wrist (scored out of 14), and Coordination/Speed (scored out of 6) for a total of 66 indicating full performance of expected motor function for the upper limb \cite{FMmethod}. Subjects that retain partial arm function and voluntary hand function, defined by an ability to grasp and release, were classified as mild (UEFM score: 38 - 66, n = 15). Subjects that (1) could not complete the Hand (/10) and Wrist (/14) sections, (2) could not display at least one finger response to upper extremity reflex tests (/4), and (3) demonstrated an inability to actively extend the paretic wrist and fingers at least 20 degrees past neutral, were classified in the severe impairment group (UEFM score: 0 - 37, n = 14). 

\subsection*{Experimental Setup}

\begin{figure*}
	\centering
		\includegraphics[scale=.215]{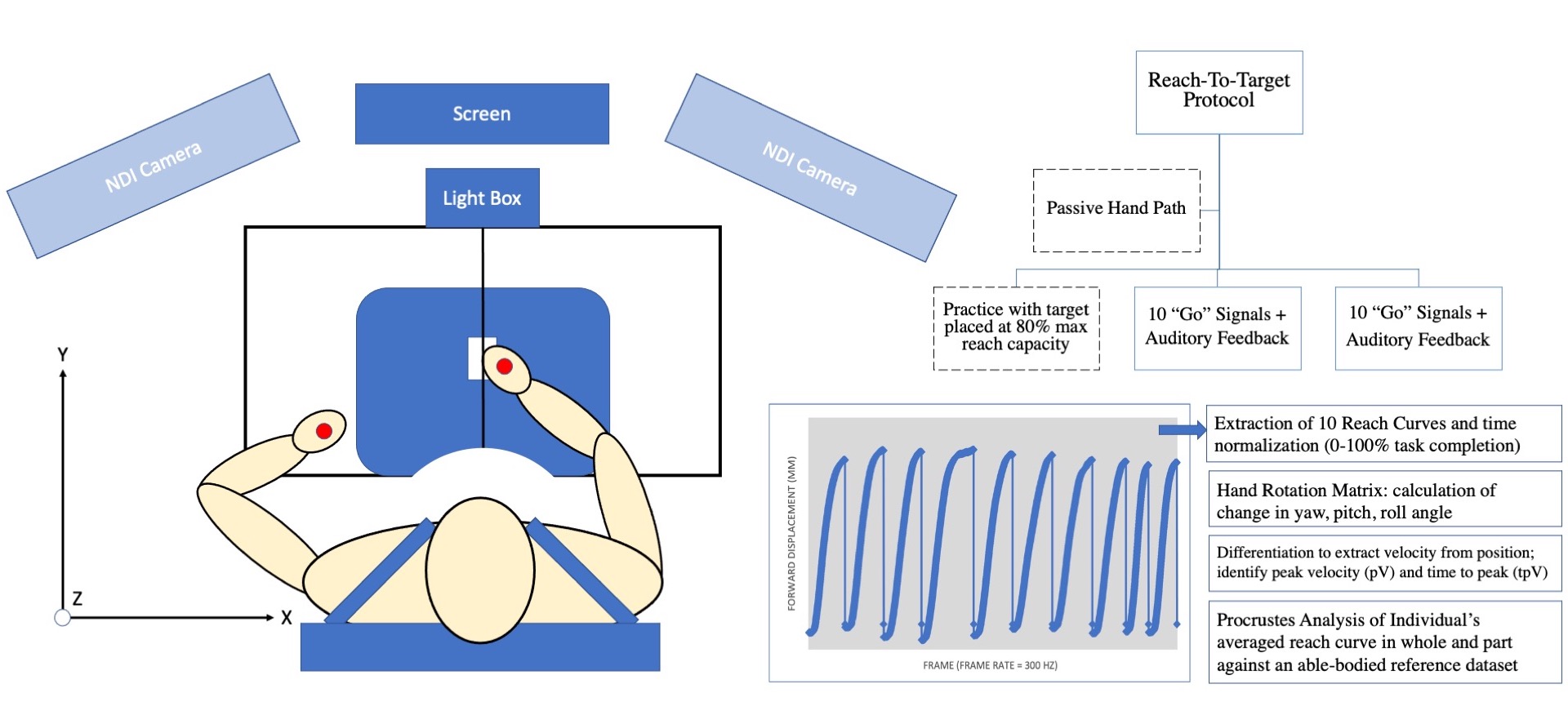}
	\caption{\textbf{Experimental Protocol and Reaching Workspace} All data collection conducted at the Mechanisms of Therapeutic Rehabilitation (MOTR) Lab at Medstar National Rehabilitation Hospital in Washington, DC. Markers placed on the hand dorsum are indicated in red. Produced 3D positional data was evaluated with custom-written MATLAB scripts to extract individual curves and kinematic metrics such as movement variability, peak velocity, time to peak velocity, and target error.}
	\label{FIG:1}
\end{figure*}

Prior to the first data collection session, subjects were familiarized with the reaching task and measurements of the chair height and distance of the chair from the table were recorded. These measurements were adjusted to ensure the subject sat as close to the table as was comfortable and maintained a 90 degree resting angle at the elbow. Subject was fitted with trunk restraints to reduce appreciable trunk involvement in the forward reaching movement \cite{Harrington20}. 

7 mm diameter IRED optical markers were placed at the dorsal surface of each hand as appropriate given each subject’s movement capability and resting hand position. A single target sensor was placed at 80\% of the maximum reach of each individual subject. Placing the target within arm's reach rather than at maximum reach capacity ensured the subject would experience typical and moderate shoulder and elbow contribution and minimize uncomfortable or compensatory movements \cite{Ma17}. Hand path kinematics were recorded using the Optotrak Certus motion capture system (Northern Digital Inc., Waterloo, Ontario, Canada) at a sampling frequency of 300 Hz, and the origin was calibrated at the front edge and center of the table at the beginning of each set of ten reaches. Optical tracking of upper extremity movement allows the collection of limb trajectory in terms of 3D Cartesian coordinates. Optotrak software was used to digitize the x-y plane in front of the subject and all movements were recording with six degree of freedom Optotrak cameras mounted surrounding and above the work-space. The relative position of the subject and the reaching workspace is depicted in Figure \ref{FIG:1}. 

Each subject completed a passive ideal hand path test in which the hand was passively moved to the target and back to represent movement without muscle activity. This measurement was used to verify and troubleshoot the collection of all positional data between the starting position and the target. Each subject completed two sets of the simple reaching test on two separate days with both the paretic and nonparetic arms. The forward reaching task was initiated after a “Go” signal was indicated either in text on a screen or a light box placed within sight of the subject. Subjects were prompted with “When the ‘Go’ signal appears, quickly reach out to touch the target” to encourage rapid forward movement. Each testing session consisted of ten “Go” signals delivered at random intervals to ensure subjects did not anticipate movement initiation.

\subsection*{Data Analysis}

Four reference curves were created from reach-to-target movements performed by two able-bodied volunteers. Able bodied persons were recruited from the Medstar Rehabilitation Hospital volunteer population. Volunteers were asked to verify they had no diagnosis of a neurological or musculoskeletal disorder that could potentially influence movement control or reaching. In order to reduce effects of hand dominance on the reference curves, one right-hand dominant and one left-hand dominant volunteer were selected. Three dimensional position data was collected from both right and left limbs first at a steady pace and then a rapid pace. The reference data set was used to create a mean healthy movement stereotype against which to analyze movements in the mild and severe impairment groups. The reference curves were compared against prior research to ensure curves were an appropriate representation of able-bodied movement. The values of the mean velocity, peak velocity, and time to peak velocity of our reference curves and values from other studies are compiled in Table \ref{tbl5}. Reference curves were used only for the Procrustes trajectory analysis; each subject's velocity, target accuracy, and orientation variability were compared between the individual's paretic and non-paretic limbs.

\begin{table*}[t]
\caption{Kinematic Metrics from Literature and Able-Bodied Reference Curves}\label{tbl5}
\begin{tabular*}{.95\linewidth}{@{\extracolsep{\fill}}llll}
\hline
Authors	& Mean Velocity (cm/s) & Peak Velocity (cm/s) & Time to Peak Velocity (\%) \\
\hline
Murphy et al \cite{Murphy11} & - & $61.6 \pm 9.4$& $46 \pm 6.9$\\
Patterson et al \cite{Patterson11} & - & $89 \pm 13$ (Comfortable) & 29.3 (single exemplar) \\
  & - & $121 \pm 14$ (Fast) & 32.8 (single exemplar) \\
Van Dokkum et al  \cite{Van14} & $42.6 \pm 5$  & 67.97 & - \\
\hline
Ref. Curve 1 & 21.24 &	57.62	& 18.20 \\
Ref. Curve 2 & 32.81	& 66.1	& 23.22 \\
Ref. Curve 3 & 23.58 &	39.72	& 26.75 \\
Ref. Curve 4 & 38.14	& 67.3	& 34.47 \\
\end{tabular*}
\newline
Reference curves 1 and 2 are left arm movements at a steady and rapid pace; curves 3 and 4 are right arm movements at a steady and rapid pace. Reference curves are used based on subject's paretic arm
\end{table*}

\begin{table*}[ht]
\caption{Kinematic metrics included in Data Analysis}\label{tbl3}
\begin{tabular*}{1\linewidth}{@{\extracolsep{\fill}}p{3.05cm}p{10cm}}
\hline
Variables & Definition used for measurements \\
\hline
Reach Duration & Time between a non-zero positive velocity followed by displacement in the positive y-direction, and a local displacement maxima which is immediately followed by a non-zero negative velocity. \\ & Normalized to 0-100\% reach completion \\
Maximum Reach & Farthest forward displacement achieved independently by subject when prompted to reach as far as they can while wearing trunk restraints \\
Mean Velocity &  The mean value during forward reach; derived from the mean velocity profile of all trials for an individual subject\\
Peak Velocity & Maximum positive velocity achieved during reach duration and corresponding to the change from acceleration to deceleration \\
Time of Peak Velocity & Percentage of total reach duration where maximum peak velocity and change from acceleration to deceleration occurs \\
Yaw Angle & $\psi$ -  Extrapolated by creating a rotation matrix A from position data every two consecutive time points, signifies the first rotation around the z-axis\\
Pitch Angle & $\theta$ - Extrapolated from rotation matrix A, rotation around the x-axis, \\
Roll Angle & $\phi$ - Extrapolated from rotation matrix A, represents the last rotation around the y-axis, i.e. the longitudinal axis of the movement arm \\
Target Error & Accuracy of the end displacement during individual reaches compared to target placed at 80\% max reach capacity \\

\end{tabular*}
\end{table*}

Individual trials of reach-to target movements were extrapolated from raw kinematic data. The beginning of a movement was defined by displacement from the starting position and a non-zero positive velocity. The completion of a movement was defined by a local maxima in position immediately followed by a non-zero negative velocity. Reach detection was confirmed by visual inspection of each trial. Two sets of consecutive reaches were averaged to create a composite curve for each individual consisting of 20 trials. The reaching trajectory data was filtered by applying a low-pass fourth order Butterworth filter with a cut-off frequency of 50 Hz to the trajectory data to account for minor variations in individual movement.

The reference trajectories were down-sampled to create ten fractions of the overall movement that were the same length as fractions of trajectories with motor impairments, in order to produce a dissimilarity profile of the overall reaching movement. Next, curve fragments composed of 35 time-points across the reference and subject curves were compared. This required a novel modified Procrustes analysis that advanced point for point along the length of the subject and reference curves to identify segments that were congruent between both. In this particular application, the curves were not scaled, since capacity to reach is specific to each subject. The index of dissimilarity, the sum of the squared Procrustes distance between each corresponding element in both curves, represents how incongruous the two segments may be, and was scaled to produce a value between 0 to 1, where 0 represents congruence between curves and 1 represents complete dissimilarity.

\subsubsection*{Kinematic Analysis}

\begin{figure*}[t]
	\centering
		\includegraphics[scale=.225]{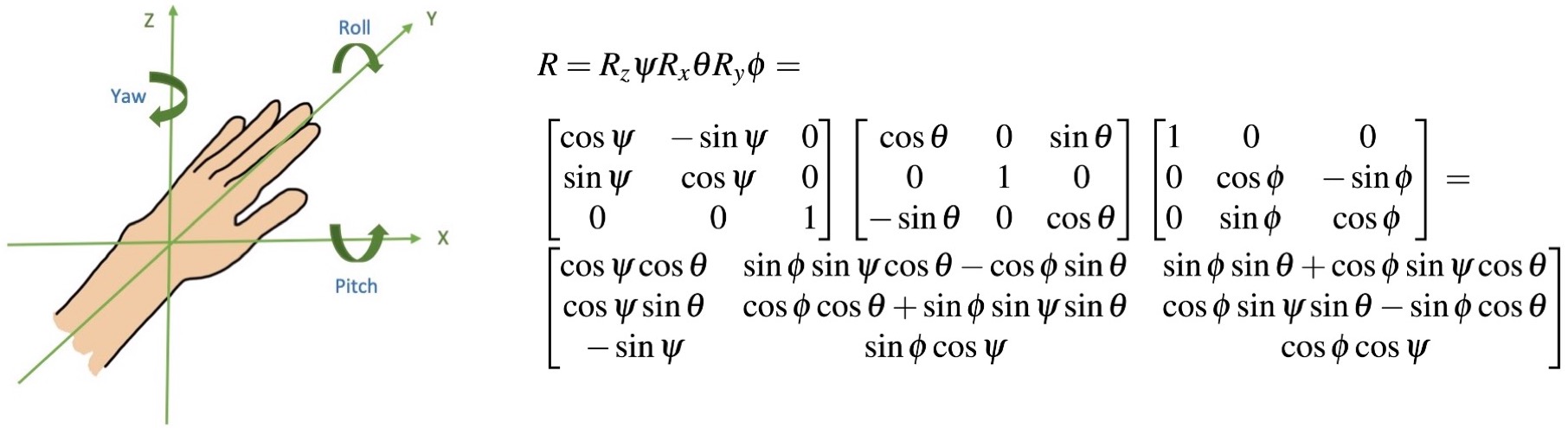}
	\caption{\textbf{Orientation of limb endpoint in 3-D space} An intrinsic coordinate system centered at the hand was used to quantify movement refinement through the reach.}
	\label{FIG:2}
\end{figure*}

The discrete kinematic metrics of interest for this study are (1) peak velocity and time to peak velocity and (2) target accuracy. The continuous metrics of interest for this study are (1) endpoint orientation and (2) curve shape. The variables were chosen to represent movement strategy and performance, as they are often reported related to outcomes of therapy. The temporal location of the discrete kinematic landmarks during reach duration was used to characterize curve shapes highlighted by the Procrustes Analysis. The captured position data were transferred to MATLAB (The MathWorks Inc) software for analysis with custom-written scripts.

For the purposes of representing online movement refinement, we utilized the recommended method of a fixed local coordinate system with respect to the work-space \cite{Bai14, Valevicius18, Wu05}. The y axis extends directly forward and represents the primary distance covered during a reaching task. The x axis extends laterally from the subject and the z axis extends inferior to superior relative to the subject (Figure \ref{FIG:2}). 

The rotations of the distal coordinate system are described in terms of the proximal coordinate system. The first rotation was described as around the z-axis, and the third rotation around the longitudinal axis, or the y-axis of the moving coordinate system. The rotation matrix in Figure \ref{FIG:2} describes the yaw-pitch-roll sequence of rotations; this was computed using consecutive data points for each incremental change in mean position during the forward reach. $\psi$ represents the yaw angle, $\theta$ represents the pitch angle, and $\phi$ represents the roll angle \cite{Sturm15}. The definitions of these angles as well as other kinematic metrics of interest are listed in Table \ref{tbl3}.

Velocity was extrapolated from the raw mean position data using the forward/backward/central differences in position data. Missing marker data were found for less than 10\% of individual trials; missing data were corrected by extrapolating from adjacent position values. A low-pass fourth order Butterworth filter with a cutoff frequency of 5 Hz was applied to the velocity data to reduce noise and distortion. The values of mean velocity, peak velocity and the time-point where peak velocity was achieved were recorded for all subject data.

Finally, each trial of forward reaching was compared to the actual location of the target as recorded for each subject. The error tolerance was adjusted to account for reaches landing within the 4 squared inches of surface area of the target pad. Positive values of target error indicate when a subject stopped movement (identified by a local maxima in displacement and subsequent movement in the negative y direction) before or at the target sensor. Negative values of target error represent when the subject has overshot or moved past the target.

\subsubsection*{Statistical Analysis}

Due to less than 50 subjects in either impairment group, an Anderson-Darling test for normalcy was performed on the kinematic metrics calculated from endpoint data \cite{NIST}. For the purposes of consistency in this paper, all statistical analyses were performed using independent t-tests and N-way ANOVA. The one-way ANOVA is mathematically equivalent to an independent t-test when applied to only two groups \cite{Aron19}. Kinematic metrics related to target error, peak velocity, and time point where peak velocity occurred were analyzed independently for differences due to impairment severity with a one-way ANOVA. Individual discrete kinematic measurements were compared in a two-way ANOVA against severity, whether the paretic limb is also the dominant limb, and which axis primarily contributed to the rotation. Separate two-way ANOVA were performed to analyze results of the modified Procrustes Analysis to interpret significance of dissimilarity indices between mild and severe impairment groups. Kinematic measurements that appeared significantly different between the mild and severe impairment groups were then used to compute preliminary RSDI scores.

\section*{Results}

All individual velocity profiles and Procrustean plots show subject exemplars from both the mild and severe impairment groups. Discrete kinematic metrics related to velocity, orientation, and target error are reported in Tables \ref{tbl6-1} and \ref{tbl6-2}, and rotation/reflection, scaling, and translation vector quantities are reported in Tables \ref{tbl7-1} and \ref{tbl7-2}.

\subsection*{Kinematic Findings}

\begin{figure*}[ht]
\centering 
\includegraphics[scale=0.12]{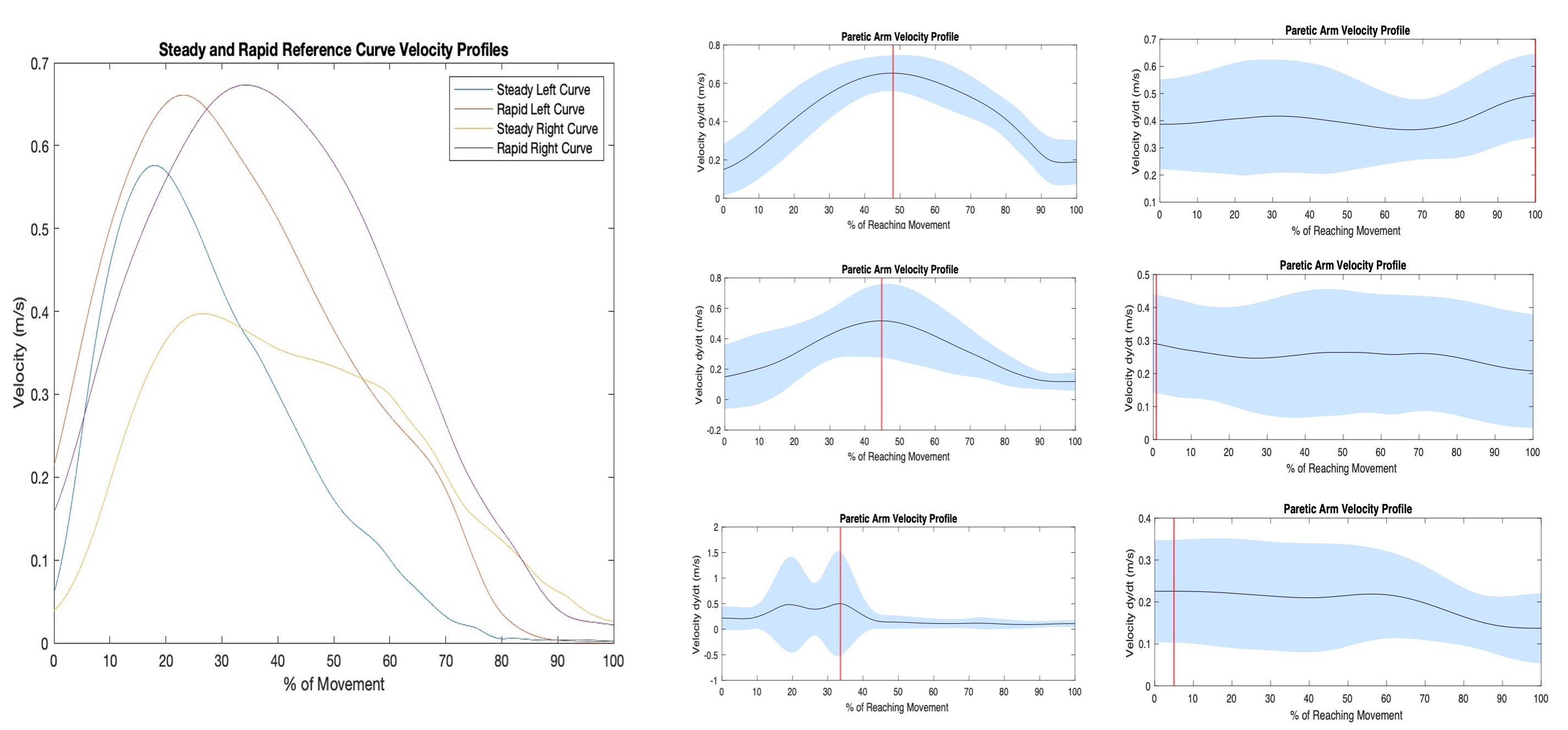} 
\caption{\textbf{Subject Exemplars of Velocity Profiles.} Left: The neural intact reference velocity curves with a steady pace and rapid pace, collected for both left-handed and right-handed movement. Middle: Three subject exemplars with mild impairment; demonstrating a delayed peak velocity. Right: Three subject exemplars with severe impairment; demonstrating peak velocities at the extremes of movement. (Note the extremely different scale of the mild impairment case in the bottom row) } 
\label{res1} 
\end{figure*}

The Anderson-Darling test for normalcy indicated that both the peak velocity [Mild: $p = 0.25$ Severe: $p = 0.73$] and velocity time location [Mild: $p = 0.30$ Severe: $p = 0.16$] were normally distributed. For subjects with mild impairment, peak velocities occurred later in the movement beyond the first third of reach progression. Some velocity profiles of subject exemplars are depicted in Figure \ref{res1}, along with the four reference curves. 

There was no significant influence of severity on subject ability to complete each set of ten reaches, nor on time needed to reach the peak velocities. However, One-Way ANOVA indicated a significant effect of group on tendency to undershoot the target ($p = 0.0214$). The severe group tended to undershoot the target with greater frequency than the mild impairment group. For subjects with severe impairment, peak velocities were lower in magnitude and occurred during movement extremes \ref{res1}. Results of statistical analyses of kinematic metrics related to accuracy and velocity are related in Table \ref{tbl4}. The Anderson-Darling test for normalcy indicated that the data collected on target error [Mild: $p = 0.69$ Severe: $p = 0.36$] was also normally distributed.

\begin{table*}[ht]
    \caption{Kinematic Findings for Mild Impairment Group}
\begin{tabular*}{.95\linewidth}{@{\extracolsep{\fill}}lllllllll}
 \hline
Sub & $V_{mean}$ (m/s) & pV (m/s) & tpV (s) & tpV (\%) & $\psi_{max}$ & $\theta_{max}$ & $\phi_{max}$ & Target Error (\%) \\
    \hline
1	&	0.33	&	0.90	&	69.43	&	0.48	&	49.89	&	99.78	&	96.31	&	$	14	\pm	2	$	\\
2	&	0.58	&	1.56	&	93.60	&	0.48	&	49.89	&	99.78	&	138.88	&	$	-24	\pm	6	$	\\
3	&	0.42	&	0.62	&	43.20	&	0.35	&	49.89	&	99.78	&	151.24	&	$	-3	\pm	3	$	\\
4	&	0.47	&	0.67	&	39.50	&	0.35	&	49.89	&	99.78	&	117.46	&	$	-10	\pm	5	$	\\
5	&	0.48	&	1.29	&	67.33	&	0.48	&	49.89	&	99.78	&	112.49	&	$	-18	\pm	3	$	\\
6	&	0.44	&	1.20	&	73.60	&	0.48	&	49.89	&	99.78	&	117.59	&	$	-2	\pm	5	$	\\
7	&	0.44	&	0.65	&	48.00	&	0.35	&	49.89	&	99.78	&	68.65	&	$	9	\pm	4	$	\\
8	&	0.38	&	0.79	&	59.60	&	0.35	&	49.89	&	99.78	&	113.78	&	$	-9	\pm	8	$	\\
9	&	0.44	&	1.87	&	84.86	&	0.35	&	49.89	&	99.78	&	151.28	&	$	-13	\pm	4	$	\\
10	&	0.31	&	0.52	&	44.80	&	0.48	&	49.89	&	99.78	&	96.13	&	$	-12	\pm	10	$	\\
11	&	0.49	&	0.85	&	54.00	&	0.48	&	49.89	&	99.78	&	122.43	&	$	-11	\pm	8	$	\\
12	&	0.42	&	0.97	&	67.60	&	0.48	&	49.89	&	99.78	&	100.69	&	$	-12	\pm	6	$	\\
13	&	0.46	&	1.25	&	70.80	&	0.48	&	49.89	&	99.78	&	109.08	&	$	-23	\pm	6	$	\\
14	&	0.22	&	0.50	&	33.60	&	0.48	&	49.89	&	99.78	&	61.26	&	$	-4	\pm	10	$	\\
15	&	0.41	&	0.83	&	66.33	&	0.48	&	49.89	&	99.78	&	111.00	&	$	-10	\pm	4	$	\\
\hline
&	0.42	&	0.97	&	61.08	&	0.44	&	49.89	&	99.78	&	111.22	&	$	-9	\pm	6	$	\\
\hline
  \end{tabular*}
  \\
\\
$V_{mean}$ - mean velocity, pV - peak velocity, tpV - time to peak velocity, maximum yaw, pitch, and roll angles, TE - Target error, negative value indicates completing short of the target
  \label{tbl6-1}
\end{table*}

\begin{table*}[ht!]
    \caption{Kinematic Findings for Severe Impairment Group}
\begin{tabular*}{.95\linewidth}{@{\extracolsep{\fill}}lllllllll}
\hline
    Sub & $V_{mean}$ (m/s) & pV (m/s) & tpV (s) & tpV (\%) & $\psi_{max}$ & $\theta_{max}$ & $\phi_{max}$ & Target Error (\%) \\
    \hline
1	&	0.25	&	0.29	&	0.83	&	0.35	&	49.89	&	99.78	&	36.15	&	$	2	\pm	18	$	\\
2	&	0.16	&	0.20	&	0.50	&	0.48	&	49.89	&	99.78	&	74.82	&	$	-15	\pm	3	$	\\
3	&	0.20	&	0.23	&	5.00	&	0.35	&	49.89	&	99.78	&	50.99	&	$	-20	\pm	12	$	\\
4	&	0.62	&	5.63	&	100.00	&	0.35	&	49.89	&	99.78	&	120.14	&	$	-90	\pm	23	$	\\
5	&	0.48	&	0.76	&	63.50	&	0.35	&	49.89	&	99.78	&	105.69	&	$	36	\pm	7	$	\\
6	&	0.26	&	0.35	&	100.00	&	0.48	&	49.89	&	99.78	&	76.39	&	$	-32	\pm	11	$	\\
7	&	0.41	&	0.73	&	71.76	&	0.35	&	49.89	&	99.78	&	120.90	&	$	-9	\pm	9	$	\\
8	&	0.41	&	0.49	&	100.00	&	0.35	&	49.89	&	99.78	&	115.16	&	$	-44	\pm	10	$	\\
9	&	0.33	&	0.40	&	42.50	&	0.48	&	49.89	&	99.78	&	99.48	&	$	-33	\pm	44	$	\\
10	&	0.32	&	0.42	&	92.00	&	0.48	&	49.89	&	99.78	&	109.38	&	$	-20	\pm	27	$	\\
11	&	0.05	&	0.06	&	2.00	&	0.48	&	49.89	&	99.78	&	55.17	&	$	-19	\pm	21	$	\\
12	&	0.44	&	0.65	&	76.00	&	0.35	&	49.89	&	99.78	&	112.56	&	$	-71	\pm	18	$	\\
13	&	0.17	&	0.30	&	0.67	&	0.35	&	49.89	&	99.78	&	60.28	&	$	-117	\pm	23	$	\\
14	&	0.09	&	0.11	&	39.00	&	0.48	&	49.89	&	99.78	&	71.70	&	$	-178	\pm	97	$	\\
\hline
	&	0.30	&	0.76	&	49.55	&	0.41	&	49.89	&	99.78	&	86.34	&	$	-43	\pm	23	$	\\
\hline
  \end{tabular*}
  \\
\\
$V_{mean}$ - mean velocity, pV - peak velocity, tpV - time to peak velocity, maximum yaw, pitch, and roll angles, TE - Target error, negative value indicates completing short of the target
  \label{tbl6-2}
\end{table*}

\begin{figure*}
\centering 
\includegraphics[scale = .2, angle=90,origin=c]{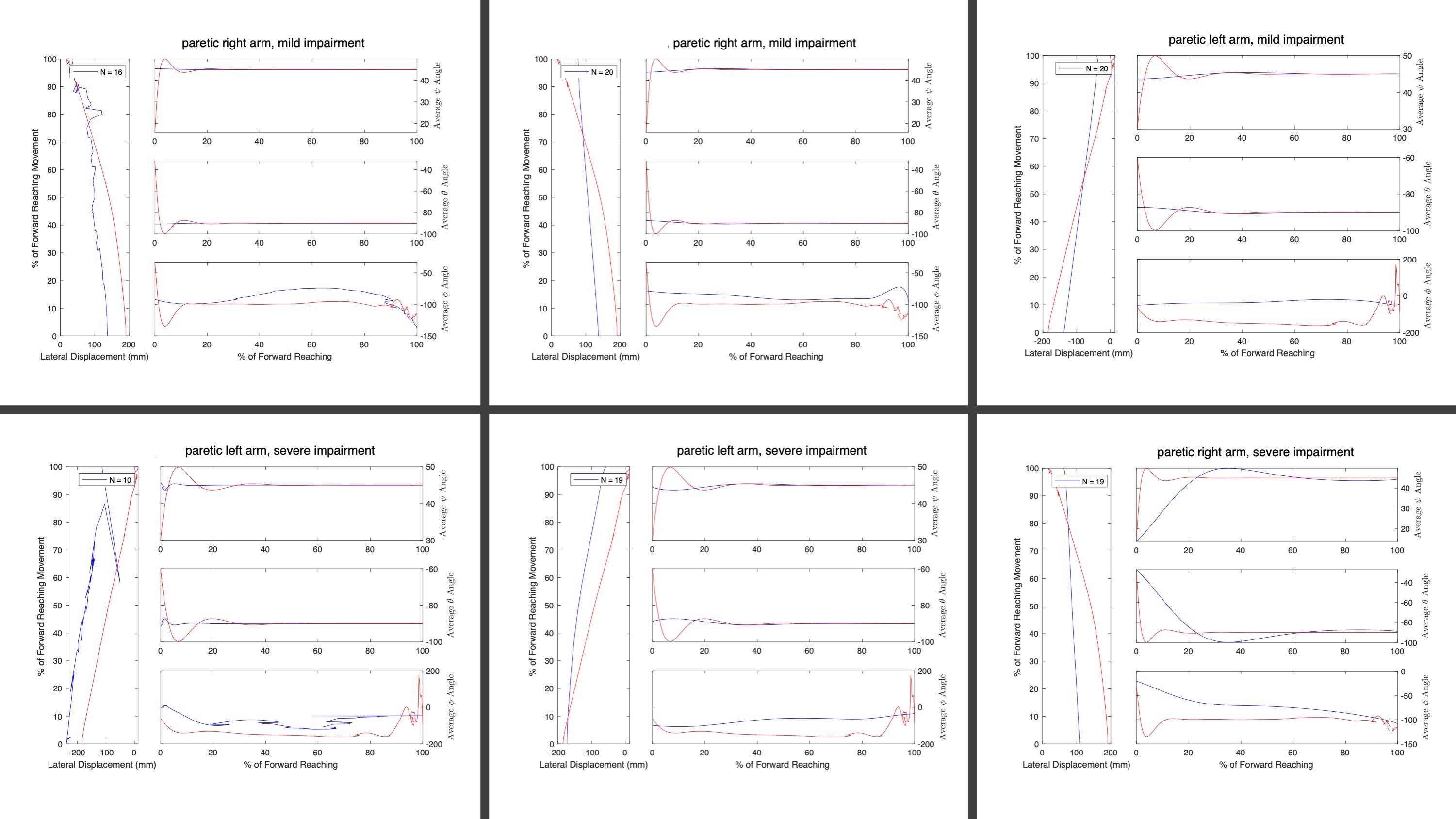} 
\caption{\textbf{Average Yaw-Pitch-Roll Angles during forward movement} Subject Exemplars from the mild and severe impairment groups of average hand orientation, where angles were extrapolated by decomposing Rotation matrices into rotation around the x-, y-, and z- axes.} 
\label{orient} 
\end{figure*}

The difference between the mild and severe impairment groups' time location of where the peak velocity occurs was analyzed with a One-Way ANOVA with a single degree of freedom, resulting in a p-value of 0.5928. In contrast, the difference between the mean velocity of the mild and severe impairment groups was found to be significant with a p-value of 0.0173. The severe impairment group tended toward more angular variability in the hand's orientation in the roll angle, or the y-axis toward movement completion. The control reach curve shows some rotation in orientation occurring in all three axes throughout the movement, as shown in Figure \ref{orient}. In contrast, in the cases of both mild and severe impairment groups, rotation could not be adequately decomposed into the yaw and pitch angles. The peak roll angles achieved during movement [Mild: 86.34, Severe: 111.22] appeared to be a significant difference between the groups, with a p-value of 0.0202. 

\begin{table*}[ht]
\caption{Analysis of Variance in Kinematic Metrics Between Mild and Severe Groups}\label{tbl4}
\begin{tabular*}{.95\linewidth}{@{\extracolsep{\fill}}llll}
\hline
Metric	&	Mild		&		Severe		&		p-value \\
\hline
Mean Velocity (m/s) & 0.42 & 0.30 & 0.0173* \\
Peak Velocity (m/s) & 0.97 & 0.76 & 0.5928 \\
Time to Peak Velocity (s) & 61.08 & 49.55 & 0.3307 \\
Time to Peak Velocity (\%) & 44 & 41 & 0.2113\\
Peak Roll Angle (deg) & 111.22 & 86.34 & 0.0202*\\
Target Error & $-9\pm6$ & $-43\pm23$ & 0.0214* \\
\\
\end{tabular*}
\end{table*}

\subsection*{Modified Procrustes Analysis Findings}

\begin{figure*}[ht]
\centering 
\includegraphics[scale=0.22]{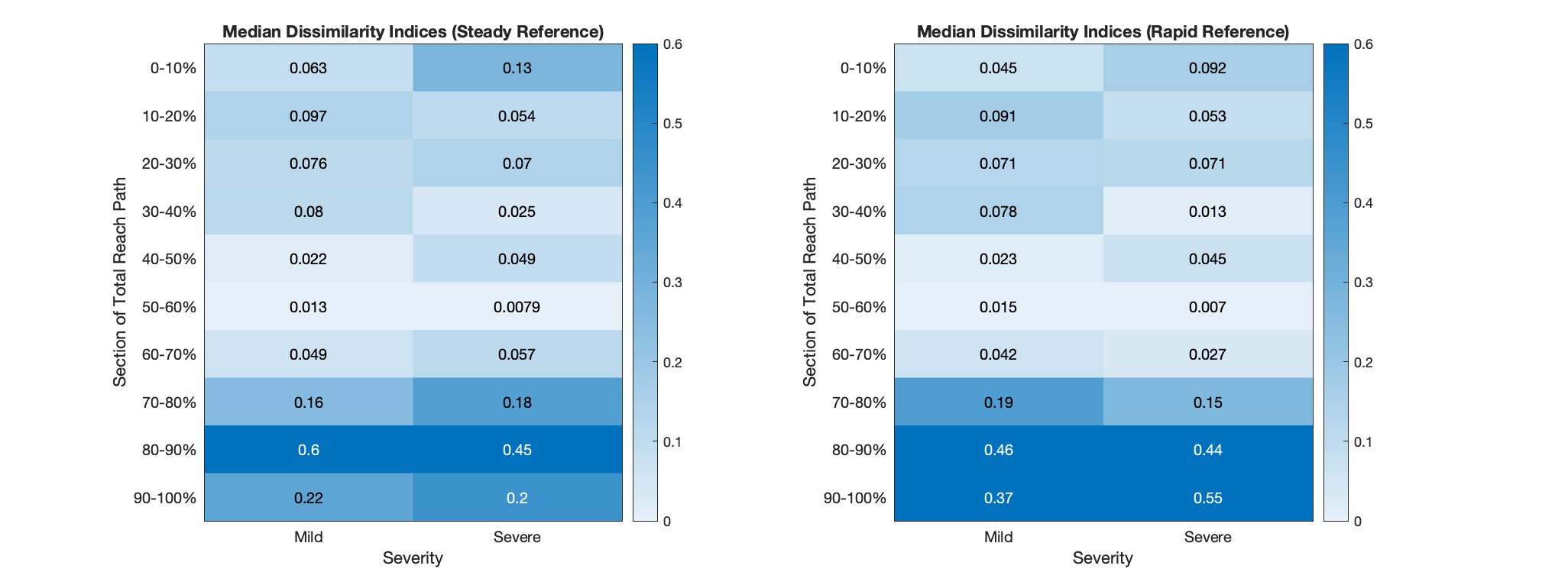} 
\caption{\textbf{Dissimilarity Indices} Heatmaps show mild and severe impairment groups compared to the steady and rapid reference curves. All groups show increased dissimilarity during movement completion, more so in the severe groups for both reference cases} 
\label{res3} 
\end{figure*}

Dissimilarity indices were calculated across ten equally sized segments of each subject curve and compared with both the steady paced and rapid paced reference curves. Visual inspection of these heatmaps indicates higher dissimilarity as movement ends in both mild and severe groups. When the groups are compared to the rapid paced reference, there is greater dissimilarity in the last three segments of movement. Though the rapid reference curve resulted from reference subjects being given the same prompt as the stroke subjects (i.e. to move as quickly as they can), this does not result in lower dissimilarities. (Figure \ref{res3}). 

\begin{table*}[ht]
    \caption{Procrustean Measures of Dissimilarity for Mild Impairment Group}
\begin{tabular*}{.95\linewidth}{@{\extracolsep{\fill}}lllllllll}

     & \multicolumn{4}{|c|}{Steady Reference Curve} & \multicolumn{4}{c}{Rapid Reference Curve} \\
    \hline
    Subject & Location (R/S) (\%) & det(T) & b & c & Loc (\%) & det(T) & b & c \\
    \hline
1	&	30	==	69	&	1	&	2.56	&	70.92	&	32	==	68	&	1	&	1.38	&	34.83	\\
2	&	3	==	51	&	1	&	11.05	&	5.39	&	6	==	56	&	-1	&	3.93	&	12.41	\\
3	&	21	==	54	&	-1	&	1.19	&	38.20	&	15	==	30	&	-1	&	0.94	&	17.06	\\
4	&	10	==	21	&	1	&	1.54	&	3.03	&	59	==	54	&	1	&	1.79	&	193.74	\\
5	&	29	==	66	&	1	&	3.79	&	100.86	&	63	==	17	&	-1	&	2.63	&	10.43	\\
6	&	7	==	63	&	1	&	11.21	&	7.05	&	66	==	12	&	1	&	3.72	&	15.00	\\
7	&	19	==	49	&	-1	&	1.53	&	19.68	&	62	==	57	&	-1	&	2.02	&	255.52	\\
8	&	8	==	46	&	-1	&	1.63	&	6.61	&	54	==	15	&	-1	&	1.21	&	12.60	\\
9	&	8	==	16	&	1	&	0.22	&	0.81	&	83	==	16	&	-1	&	2.46	&	21.70	\\
10	&	50	==	45	&	1	&	1.56	&	111.87	&	49	==	53	&	1	&	0.94	&	51.20	\\
11	&	57	==	46	&	-1	&	1.99	&	203.81	&	39	==	19	&	1	&	1.49	&	7.08	\\
12	&	31	==	65	&	-1	&	3.12	&	103.69	&	61	==	19	&	1	&	2.11	&	6.59	\\
13	&	10	==	63	&	-1	&	9.33	&	6.46	&	63	==	10	&	1	&	4.02	&	11.50	\\
14	&	66	==	45	&	1	&	0.54	&	26.66	&	62	==	14	&	1	&	0.34	&	75.37	\\
15	&	30	==	67	&	1	&	3.02	&	67.22	&	67	==	33	&	1	&	1.62	&	23.56	\\
\hline
 &  &  & 3.62  & 51.48 &  &  & 2.04 & 49.91 \\
\hline
  \end{tabular*} \label{tbl7-1}
\end{table*}

\begin{table*}[ht!]
    \caption{Procrustean Measures of Dissimilarity for Severe Impairment Group}
\begin{tabular*}{.95\linewidth}{@{\extracolsep{\fill}}lllllllll}

     & \multicolumn{4}{|c|}{Steady Reference Curve} & \multicolumn{4}{c}{Rapid Reference Curve} \\
    \hline
    Subject & Location (R/S) (\%) & det(T) & b & c & Loc (\%) & det(T) & b & c \\
\hline
1	&	12	==	60	&	1	&	1.19	&	23.85	&	13	==	58	&	-1	&	0.97	&	27.86	\\
2	&	57	==	24	&	1	&	1.38	&	160.57	&	20	==	58	&	1	&	0.75	&	57.74	\\
3	&	92	==	66	&	-1	&	9.37	&	2029.10	&	28	==	65	&	1	&	0.78	&	17.75	\\
4	&	10	==	38	&	1	&	0.24	&	5.83	&	12	==	38	&	1	&	0.19	&	6.56	\\
5	&	13	==	60	&	1	&	1.76	&	13.00	&	15	==	61	&	-1	&	1.43	&	25.10	\\
6	&	11	==	57	&	-1	&	2.86	&	28.01	&	55	==	8	&	1	&	1.58	&	28.20	\\
7	&	13	==	71	&	1	&	1.84	&	11.00	&	15	==	72	&	-1	&	1.5	&	19.05	\\
8	&	20	==	39	&	-1	&	1.18	&	36.28	&	27	==	22	&	-1	&	0.88	&	21.37	\\
9	&	17	==	26	&	-1	&	3.83	&	43.65	&	43	==	43	&	-1	&	1.58	&	112.37	\\
10	&	59	==	19	&	1	&	1.80	&	245.35	&	19	==	53	&	-1	&	1.00	&	120.51	\\
11	&	64	==	37	&	-1	&	4.71	&	716.18	&	37	==	52	&	-1	&	2.33	&	274.91	\\
12	&	5	==	15	&	1	&	1.12	&	0.54	&	83	==	67	&	-1	&	3.01	&	512.38	\\
13	&	22	==	63	&	1	&	0.77	&	34.93	&	64	==	36	&	1	&	0.61	&	31.34	\\
14	&	18	==	42	&	-1	&	2.83	&	21.83	&	49	==	40	&	-1	&	1.20	&	65.56	\\
\hline
 &  &  & 2.49 & 240.72 &  & & 1.27 & 94.33\\
\hline
  \end{tabular*}
  \\
\\
Center locations of reference and subject curve segments that are congruent, det(T) - Determinant of the Reflection/Rotation matrix, where 1 indicates a rotation and -1 indicates a reflection , b - Scaling component, c - Magnitude of translation vector to conform subject curve to the reference
  \label{tbl7-2}
\end{table*}

\begin{table*}[!ht]
\caption{Analysis of Procrustes Transformation Variables Between Mild and Severe Groups}\label{tbl8}
\begin{tabular*}{.95\linewidth}{@{\extracolsep{\fill}}l|lll|lll}

&   & Steady Reference  &  &  & Rapid Reference   & \\
\hline
	&	Mild		&		Severe		&		p-value &	Mild		&		Severe		&		p-value \\
\hline
Reference Segment &&& 0.6288 &&& 0.0329* \\
Subject Segment &&& 0.287 &&& 0.0285* \\
Rotation/Reflection &&& 0.8813 &&& 0.2043 \\
Scaling & 3.62 & 2.49 & 0.3406 & 2.04 & 1.27 & 0.0397*\\
Translation & 51.48 & 240.72 & 0.1947 & 49.91 & 94.33 & 0.2889 \\
\end{tabular*}
\end{table*}

\begin{table*}[!ht]
\caption{Analysis of Variance in Severity and Congruence to Reference. Constrained (Type III) Sum of Squares. }\label{res4}
\begin{tabular*}{.95\linewidth}{@{\extracolsep{\fill}}llll}
\hline
Source	&	Sum Sq		&		d.f.		&		p-value \\
\hline
Severity (Mild/Severe)	&	31.1	&	1	&	0.4656	\\
Steady Ref Movement & 5034.8 & 1 & 0*\\
Sub Movement & 481 & 1 & 0.0096*\\
Severity*Ref Movement & 120.6 & 2 & 0.3619\\
Severity*Sub Movement & 28.9&2&0.7747\\
Ref*Sub & 354 & 3 & 0.1378\\
\hline
Severity (Mild/Severe)	&	179.8	&	1	&	0.2059	\\
Rapid Ref Movement & 2671.4 & 1 & 0.0001*\\
Sub Movement & 2826.8 & 1 & 0.0001*\\
Severity*Ref Movement & 227.8 & 2 & 0.3564\\
Severity*Sub Movement & 182.8 &2&0.4324\\
Ref*Sub & 293.6 & 3 & 0.4415\\
\end{tabular*}
\end{table*}

\begin{table*}[!ht]
\caption{Analysis of Variance in Arm Dominance and Severity. Constrained (Type III) Sum of Squares. }\label{res5}
\begin{tabular*}{.95\linewidth}{@{\extracolsep{\fill}}llll}
\hline
Source	&	Sum Sq		&		d.f.		&		p-value \\
\hline
Severity (Mild/Severe)	&	1732.7	&	1	&	0.0536*	\\
Dominance of Paretic & 3277.3 & 1 & 0.0107*\\
Sub differences from Steady Ref & 5711.4 & 2 & 0.0053*\\
Severity*Dominance & 279.8 & 1 & 0.4186\\
Severity*Sub Movement & 428.1 & 2 & 0.601\\
Dominance*Sub & 134.2 & 2 & 0.85\\
\hline
Severity (Mild/Severe)	&	36.1 & 1 & 0.7815	\\
Dominance of Paretic & 82.6 & 1 & 0.6753 \\
Sub differences from Rapid Ref & 14635.2 & 2 & 0.0001*\\
Severity*Dominance & 1188.4 & 1 & 0.1231 \\
Severity*Sub Movement & 981.5 & 2 & 0.3611\\
Dominance*Sub & 709.3 & 2 & 0.4738\\
\end{tabular*}
\end{table*}

\begin{figure*}
\centering 
\includegraphics[scale = 0.2, angle=90,origin=c]{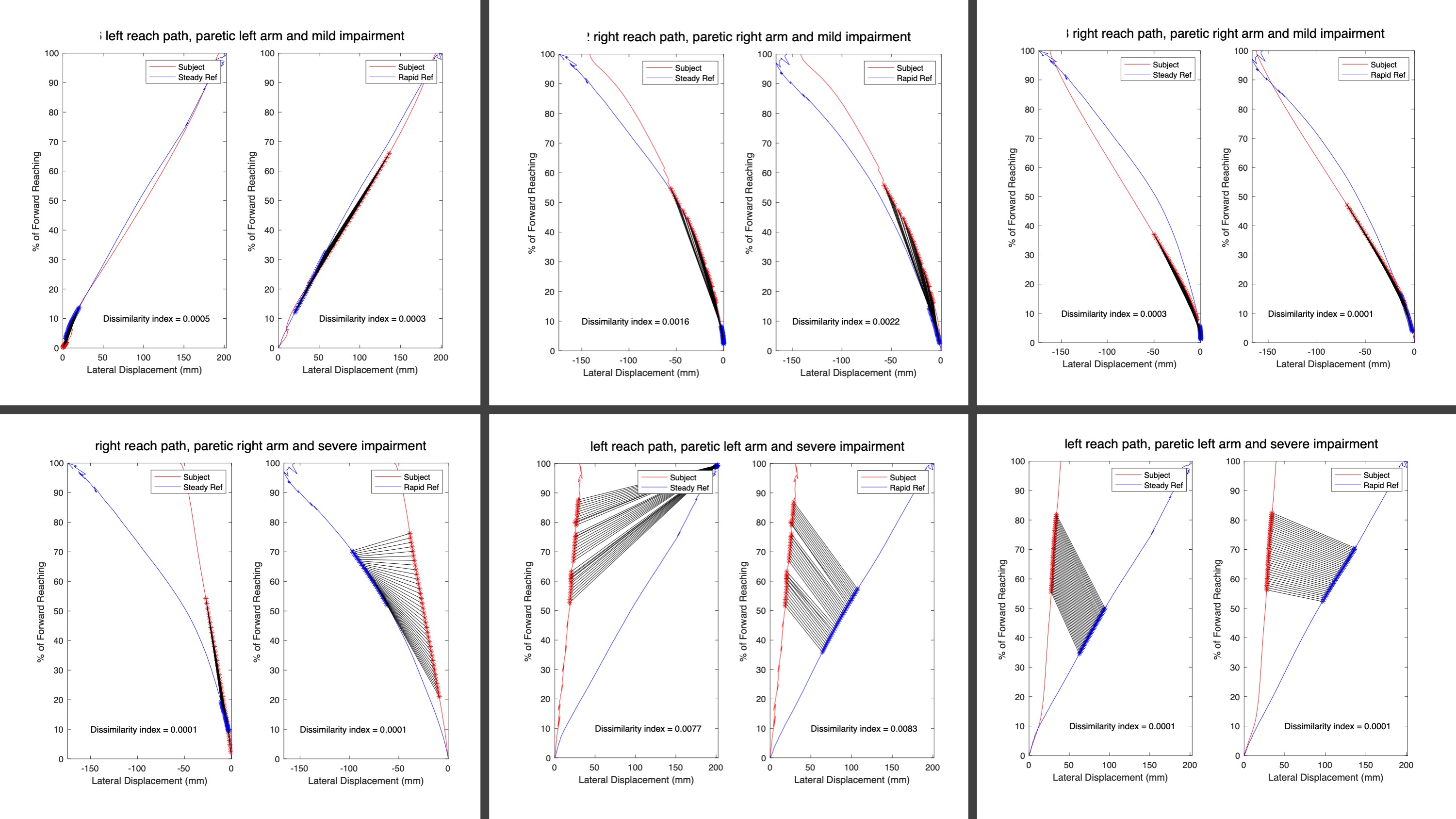} 
\caption{\textbf{Sub-movements in the mild impairment group remain congruous to healthy movement.} Modified Procrustes Analysis of individuals with mild impairment shows movement initiation proceeds similarly to the steady and rapid movement curves. Individuals with severe impairment do not have meaningful congruous behaviors to reference movement.} 
\label{res2} 
\end{figure*}

When the complete subject reach path was compared to the complete control reach path with a One-Way ANOVA, there was no significant effect of severity on curve dissimilarity between the mild and severe groups ($p=0.62$). The Procrustes Method was then modified to compare segments, defined as 35 consecutive time-points, by advancing along the mean individual and control reach curve point for point. In all cases of mild impairment, some of which are depicted in Figure \ref{res2}, the initial subject kinematic behavior appears most congruous to the initial control kinematic behavior. Regardless of overall response time, in both two- and three-dimensional representations of movement, movement initiation proceeds comparably to the healthy control curve. The modified Procrustes analysis showed the initial impulse control phase to be evident and preserved in stroke survivors with mild functional impairment but not with severe impairment. The portion of movement in the mild impairment group that replicated the control movement not only occurred in the initial phase of movement, but also occurred before the peak velocity was achieved.

Table \ref{tbl8} details the analysis of variance in the rotation, scaling, and translation transformation variables found through Procrustes analysis of the most congruent subject and reference segments. The mean scaling factors when compared to the smooth reference curve [Mild: 3.62, Severe: 2.49], and rapid reference curve [Mild: 2.04, Severe: 1.27] all indicate that the impairment groups demonstrated stretched movement, i.e. the subjects  took longer amounts of time than the reference to complete the specific segment of movement. Subjects demonstrated an ability to prioritize and modulate speed of movement by decreasing the time required to complete the specific segment of movement. The difference between the mild and severe impairment groups produced a p-value of 0.0397. 

N-Way ANOVA tests were performed to analyze the influence of severity and dominance on the time-location of the congruent segments in the subject and reference, and on the time-duration of the subject movement that appeared congruent to the reference. The complete analysis of the main effects and interaction effects on the location of congruent subject and reference segments is detailed in Tables \ref{res4} and \ref{res5}. A three-way ANOVA was performed assessing the significance of dissimilarity indices of the following factors: impairment severity, the location in the subject behavior where the curve dissimilarity occurs, and the location of the control behavior that is most likely preserved in subject behavior. Where there were no significance of the main effects, the two-way interaction of each of the three factors showed significance, as detailed in Table \ref{res5}. Severity and the preservation of movement initiation do not show significant interaction effects ($p = 0.4656$). While there was no independent effect of the paretic limb also being the dominant limb ($p = 0.6753$) when the rapid reference curve is used for comparison, hand dominance contributes to a significant difference between populations when the reference motivation is to produce steady and smooth movement, ($p = 0.0107$). . The population marginal means of the groups of mild impairment with a paretic non-dominant limb, and severe impairment with a paretic non-dominant limb are significantly different. The population marginal means for both groups of impairment where the paretic limb is the dominant limb did not have any significant differences. 

The analysis of the main effects and interaction effects on the length of the congruent subject segment is detailed in Table \ref{res6}. A three-way ANOVA was performed assessing the significance of dissimilarity indices of the following factors: impairment severity, the location in the subject behavior where the curve dissimilarity occurs, and whether the paretic arm was also the dominant arm. The impairment severity classification of the subject had a significant main effect on the length of congruence of the subject segment, p-value of 0.0342. Where there were no significance of the other main effects, the two-way interaction of severity and arm dominance had a p-value of 0.0364. 

\begin{table*}[ht]
\caption{Analysis of Variance in Influence of Arm Dominance and Severity on Size of Congruent Segments. Constrained (Type III) Sum of Squares. }\label{res6}
\begin{tabular*}{.95\linewidth}{@{\extracolsep{\fill}}llll}
\hline
Source	&	Sum Sq		&		d.f.		&		p-value \\
\hline
Severity (Mild/Severe)	&	1072.27	&	1	&	0.0342*	\\
Dominance of Paretic & 393.52 & 1 & 0.1828\\
Subject Movement & 965.76 & 2 & 0.1229\\
Severity*Dominance & 1042.78 & 1 & 0.0364*\\
Severity*Sub Movement & 180.53 & 2 & 0.6514\\
Dominance*Sub & 56.9 & 2 & 0.8718\\

\end{tabular*}
\end{table*}

\subsection*{Preliminary severity and dissimilarity scores}

Metrics related to kinematics and the modified Procrustes analysis that showed significant differences between the mild and severe populations were used to compute RSDI-Severity and RSDI-Dissimilarity sub-scores. The severity sub-score comprised of velocity, orientation, and accuracy elements while the dissimilarity sub-score comprised of dissimilarity indices of the overall movement, the ending movements, and the location and length of the reference segment that was found to be most congruous. The scaling components were also included in the computation of the dissimilarity sub-score.

The preliminary RSDI sub-scores computed using these metrics were classified in terms of likely rehabilitation goals. Subjects with a higher severity indices and lower dissimilarity indices due to low mean velocities, low peak angular values, and high target error, may benefit from a classification that prioritizes speed-focused goals. Such subjects were given a "Speed Emphasis" classification. Alternatively, subjects with lower severity indices and higher dissimilarity indices were scored as such due to high dissimilarity to the reference movement, or elongated movement behaviors, implying a need for "Strength Emphasis" to produce stable movements. Subjects with comparable severity and dissimilarity indices were classified as "Combined Emphasis". These classifications compared with the UEFM mild and severe classifications are cross tabulated in Table \ref{RSDI}.

\begin{table*}[ht]
\caption{Cross Tabulation Table for Upper Extremity Fugl-Meyer (UEFM) and Reach Severity \& Dissimilarity Index (RSDI)}\label{RSDI}
\begin{tabular*}{.95\linewidth}{@{\extracolsep{\fill}}llll}

 & \multicolumn{3}{|c|}{RSDI} \\
\hline
& Speed Emphasis & Strength Emphasis & Combined Emphasis \\
\hline
Mild & 0 & 10 & 5\\
Severe & 5 & 7 & 2\\

\end{tabular*}
\end{table*}

The first row in Table \ref{RSDI} shows that of the 15 subjects classified as mildly impaired according to the UEFM test, 10 received a Strength Emphasis and 5 received a Combined emphasis. This is consistent with the clinical observation that persons with mild impairment continue to be able to reach forward quickly while compensating for muscle weakness and loss of agility. The second row indicates that of the 14 subjects classified as severely impaired by the UEFM test, 5 can be reclassified as Speed Emphasis, 7 as Strength Emphasis, and 2 as Combined Emphasis.

\section*{Discussion}

During the reach to target movement performed in this study, the hand passes medially to reach the target which is centered in front of the subject, in addition to forward displacement. The position of the hand as it moves through space was captured as endpoint data representing the movement of the arm. We can extrapolate kinematic metrics such as mean velocity, peak velocity, the time required to achieve peak velocity, and target accuracy from this endpoint data. Additionally, the position of the hand over time can be compared to reference datasets in order to quantify deviation of the arm during a reach to target movement.  

 The subjects classified as mildly impaired in this study achieved higher mean velocities than their severely impaired counterparts. Another quantity found to differ significantly between impairment groups was the target accuracy. Higher target error may be correlated with diminished ability to sub-correct movements during the final phase of movement where precision and accuracy is prioritized. Earlier motor control decision-making prioritizes speed and minimization principles. The data thus lends some support to the observation that response time and target accuracy are disrupted after stroke but not physical capability of ballistic movement. Movement is modulated differently during reaching, with every particular functional limitation requiring an investigation of which kinematic metrics require incorporation into deciding the best therapeutic interventions.

We found the range of roll angles achieved by the arm to also differ significantly between impairment groups. Mildly impairment individuals demonstrated higher peak roll angles, whereas individuals with severe impairments had much lower rotation around the y-axis to achieve a lateral-to-medial movement in front of the subject. This could potentially imply a phase of movement where movement is constrained by maladaptive joint movement, such as a compensatory adaptation between the elbow and shoulder, with joint movement becoming inflexible during the forward movement. This may be due to range of motion being constrained while speed is prioritized over accuracy or online movement correction. Clinically, these findings could translate to the development of tasks where the target is placed elsewhere in the three-dimensional space in front of the subject for more effective reaching practice, e.g. a ball suspended in the air, targets placed radially equidistant, etc. A particular subject may need to be motivated not by reaction time, but by following a pre-drawn path as precisely as possible.

 The dissimilarity indices of specific events with the reaching task are of particular interest, and imply that some movement behavior is preserved in mild impairment that is disrupted with severe impairment. A most interesting finding of the modified Procrustes analysis is that severity has a significant interaction effect, along with hand dominance, on whether a subject replicates reference behavior while initiating reach or at some point during the reach task. Individuals with mild impairments replicated reference behavior when beginning movement.The relative timing of the peak velocity within the first phase of movement follows prior literature describing the initiation of movement being based on anticipation of the task and not sensory feedback. Applying dissimilarity indices to the overall movement may represent an overall effect of impairment severity. The modified Procrustes method, alternatively, allowed dissimilarity indices to be computed across segments of the entire movement. Both subjects with mild and severe impairment showed that completion movements were not similar to the reference data, though they deviated more from the reference in the case of severe impairment. In the clinical setting, a subject demonstrating congruous movement initiation may focus on precision exercises and visual feedback incorporation, while a subject demonstrating congruous movement completion may practice speed exercises and need not emphasize target accuracy.

\section*{Conclusions}

While rehabilitation efforts can be effectively informed by clinical observation in the case of individuals with mild functional impairments, individuals exhibiting severe impairments require a deeper investigation of when and how deficits emerge. The tri-phasic activation pattern of upper extremity movement and the behavioral model of rapid movement, error correction, and precision control imply that movement may be disrupted in different ways in different parts of the reach-to-target task.  The use of endpoint kinematic data does not allow for decomposition of rotation matrices to identify specific joint contributions; however, it can be used to identify differences in velocity, accuracy, smoothness, and deviation from reference movements. Though the upper extremity is neither cyclical nor stereotyped in its movement like the lower extremity, nevertheless measurements of gait deviation can guide analogous measures of severity and dissimilarity for the arm during functional sub-movements such as the reach and grasp cycle. 

The Modified Procrustes method produced intriguing results that are supported by clinical observations; namely that mild impairment does not exhibit a disruption in the ability to initiate rapid movement. By comparing curved paths point by point, clinicians may pinpoint when a disruption in movement occurs. Taking into account how the overall limb is oriented when this disruption occurs could then allow for only specific joint measurements to be taken rather than throughout the movement. This creates the possibility for movement tracking to remain simple yet effective, so that it can be incorporated into the clinical setting without increasing patient burden.

The RSDI score proposed in this paper can be applied to any patient position data, provided the clinician also has access to reference datasets. The RSDI can thus also be expanded to other movements, if such movements have also been recorded by healthy volunteers. In this way, the  RSDI score can easily be adapted and modified to a given clinician's protocol, and provide insight when creating rehabilitation goals. It would also be worthwhile to expand the methods explored in this paper to multi-joint models of the arm to objectively identify the presence of synergies or compensatory movements that may then be incorporated into rehabilitative practice. Collecting multi-joint data by centering visual markers on each limb segment will allow for characterization of joint contributions to movement deficits. Although the RSDI preliminary results only include a few metrics of upper extremity movement, we hope in future studies it can continue to be refined and expanded to include other functional movements. This study did not accommodate for differences in limb dominance, an important consideration for future studies as limb dominance certainly has an impact on rehabilitation and quality of life. Another limitation of the current study 

The upper extremity presents a rich platform for studying the motor system and how it is affected by the physical world around it and the internal world that controls and communicates through it. Through advancing the kinematic questions explored in this study and understanding the specific control parameters and factors that constrain and alter function, we hope that the impairment and functional limitations correlated with stroke may be minimized and thus prevented from translating to disability in social functioning. By creating a comprehensive and objective clinical tool to select rehabilitative strategies that can serve each individual's specific needs, we anticipate the impact of stroke on disability and quality of life may be appreciably reduced. 

\section*{Acknowledgments}

We are indebted to all who participated in this study. We would like to thank Drs. Rachael Harrington and Evan Chan for guidance with experiment design and assistance with data collection, Dr. Kathryn Laskey for guiding and reviewing the statistical analyses, and Dr. Qi Wei and Dr. Joseph Majdi for editing and reviewing this paper. Finally, we are grateful to the MedStar National Rehabilitation Hospital research department for facilitating recruitment for this project.


\begin{thebibliography}{99}
\bibitem{Andel08}
C. J. van Andel et al,
\emph{Complete 3D kinematics of upper extremity functional tasks},
\emph{Gait and Posture},
vol. 27, no. 1, pp. 120--127, Jan. 2008.

\bibitem{Cai19}
L. Cai et al,
\emph{Validity and Reliability of Upper Limb Functional Assessment Using the Microsoft Kinect V2 Sensor},
\emph{Applied Bionics and Biomechanics},
vol. 2019, pp. 1--14, Feb. 2019.

\bibitem{Collins18}
K. C. Collins et al,
\emph{Kinematic Components of the Reach-to-Target Movement After Stroke for Focused Rehabilitation Interventions: Systematic Review and Meta-Analysis},
\emph{Frontiers in Neurology},
vol. 9, Jun. 2018.
doi: 10.3389/fneur.2018.00472

\bibitem{Cooke90}
J. D. Cooke and S. H. Brown,
\emph{Movement-related phasic muscle activation. {II}. Generation and functional role of the triphasic pattern},
\emph{Journal of Neurophysiology},
vol. 63, no. 3, pp. 465--472, Mar. 1990.
doi: 10.1152/jn.1990.63.3.465

\bibitem{Elliott20}
D. Elliott et al,
\emph{The multiple process model of goal-directed aiming/reaching: insights on limb control from various special populations},
\emph{Experimental Brain Research},
vol. 238, no. 12, pp. 2685--2699, Oct. 2020.
doi: 10.1007/s00221-020-05952-2

\bibitem{Elliott10}
D. Elliott et al,
\emph{Goal-directed aiming: Two components but multiple processes},
\emph{Psychological Bulletin},
vol. 136, no. 6, pp. 1023--1044, 2010.
doi: 10.1037/a0020958

\bibitem{Elliott01}
D. Elliott et al,
\emph{A century later: Woodworth's (1899) two-component model of goal-directed aiming},
\emph{Psychological Bulletin},
vol. 127, no. 3, pp. 342--357, 2001.
doi: 10.1037/0033-2909.127.3.342

\bibitem{HarrisLove12}
M. Harris-Love,
\emph{Transcranial Magnetic Stimulation for the Prediction and Enhancement of Rehabilitation Treatment Effects},
\emph{Journal of Neurologic Physical Therapy},
vol. 36, no. 2, pp. 87--93, Jun. 2012.
doi: 10.1097/npt.0b013e3182564d26

\bibitem{JarqueBou20}
N. J. Jarque-Bou et al,
\emph{Hand Kinematics Characterization While Performing Activities of Daily Living Through Kinematics Reduction},
\emph{{IEEE} Transactions on Neural Systems and Rehabilitation Engineering},
vol. 28, no. 7, pp. 1556--1565, Jul. 2020.
doi: 10.1109/tnsre.2020.2998642

\bibitem{Jaspers11}
E. Jaspers et al,
\emph{Upper limb kinematics: Development and reliability of a clinical protocol for children},
\emph{Gait and Posture},
vol. 33, no. 2, pp. 279--285, Feb. 2011.
doi: 10.1016/j.gaitpost.2010.11.021

\bibitem{Jaspers11APS}
E. Jaspers et al,
\emph{The Arm Profile Score: A new summary index to assess upper limb movement pathology},
\emph{Gait \& Posture},
vol. 34, no. 2, pp. 227--233, 2011.

\bibitem{Krauth19}
R. Krauth et al,
\emph{Cortico-Muscular Coherence Is Reduced Acutely Post-stroke and Increases Bilaterally During Motor Recovery: A Pilot Study},
\emph{Frontiers in Neurology},
vol. 10, Feb. 2019.
doi: 10.3389/fneur.2019.00126

\bibitem{Mesquita20}
I. A. Mesquita et al,
\emph{Comparison of upper limb kinematics in two activities of daily living with different handling requirements},
\emph{Human Movement Science},
vol. 72, pp. 102632, Aug. 2020.
doi: 10.1016/j.humov.2020.102632

\bibitem{Mesquita19_Part2}
I. A. Mesquita et al,
\emph{Methodological considerations for kinematic analysis of upper limbs in healthy and poststroke adults Part II: a systematic review of motion capture systems and kinematic metrics},
\emph{Topics in Stroke Rehabilitation},
vol. 26, no. 6, pp. 464--472, May 2019.
doi: 10.1080/10749357.2019.1611221

\bibitem{Mesquita19_Part1}
I. A. Mesquita et al,
\emph{Methodological considerations for kinematic analysis of upper limbs in healthy and poststroke adults. Part I: A systematic review of sampling and motor tasks},
\emph{Topics in Stroke Rehabilitation},
vol. 26, no. 2, pp. 142--152, Nov. 2018.
doi: 10.1080/10749357.2018.1551953

\bibitem{Morel17}
P. Morel et al,
\emph{What makes a reach movement effortful? Physical effort discounting supports common minimization principles in decision making and motor control},
\emph{{PLOS} Biology},
vol. 15, no. 6, p. e2001323, Jun. 2017.
doi: 10.1371/journal.pbio.2001323

\bibitem{Szwedo21}
E. N. Szwedo et al,
\emph{Test-retest repeatability reveals a temporal kinematic signature for an upper limb precision grasping task in adults},
\emph{Human Movement Science},
vol. 75, p. 102721, Feb. 2021.
doi: 10.1016/j.humov.2020.102721

\bibitem{Ozturk16}
A. Ozturk et al,
\emph{A clinically feasible kinematic assessment method of upper extremity motor function impairment after stroke},
\emph{Measurement},
vol. 80, pp. 207--216, Feb. 2016.
doi: 10.1016/j.measurement.2015.11.026

\bibitem{Poitras19}
I. Poitras et al,
\emph{Validity and Reliability of Wearable Sensors for Joint Angle Estimation: A Systematic Review},
\emph{Sensors},
vol. 19, no. 7, p. 1555, Mar. 2019.
doi: 10.3390/s19071555

\bibitem{Priot20}
A.-E. Priot et al,
\emph{Sensory Prediction of Limb Movement Is Critical for Automatic Online Control},
\emph{Frontiers in Human Neuroscience},
vol. 14, Oct. 2020.
doi: 10.3389/fnhum.2020.549537

\bibitem{Reissner19}
L. Reissner et al,
\emph{Assessment of hand function during activities of daily living using motion tracking cameras: A systematic review},
\emph{Proceedings of the Institution of Mechanical Engineers, Part H: Journal of Engineering in Medicine},
vol. 233, no. 8, pp. 764--783, May 2019.
doi: 10.1177/0954411919851302

\bibitem{Guzman14}
A. de los Reyes-Guzm{\'{a}}n et al,
\emph{Quantitative assessment based on kinematic measures of functional impairments during upper extremity movements: A review},
\emph{Clinical Biomechanics},
vol. 29, no. 7, pp. 719--727, Aug. 2014.
doi: 10.1016/j.clinbiomech.2014.06.013

\bibitem{Roberts20}
J. W. Roberts,
\emph{Energy minimization within target-directed aiming: the mediating influence of the number of movements and target size},
\emph{Experimental Brain Research},
vol. 238, no. 3, pp. 741--749, Feb. 2020.
doi: 10.1007/s00221-020-05750-w

\bibitem{Roberts19}
J. W. Roberts and G. P. Lawrence,
\emph{Impact of attentional focus on motor performance within the context of "early" limb regulation and "late" target control},
\emph{Acta Psychologica},
vol. 198, pp. 102864, Jul. 2019.
doi: 10.1016/j.actpsy.2019.102864

\bibitem{Saes22}
M. Saes et al,
\emph{Quantifying quality of reaching movements longitudinally post-stroke: a systematic review},
\emph{Neurorehabilitation and neural repair},
vol. 36, no. 3, pp. 183--207, 2022.

\bibitem{Scano19}
A. Scano et al,
\emph{Low-Cost Tracking Systems Allow Fine Biomechanical Evaluation of Upper-Limb Daily-Life Gestures in Healthy People and Post-Stroke Patients},
\emph{Sensors},
vol. 19, no. 5, pp. 1224, Mar. 2019.
doi: 10.3390/s19051224

\bibitem{Schwartz16}
A. B. Schwartz,
\emph{Movement: How the Brain Communicates with the World},
\emph{Cell},
vol. 164, no. 6, pp. 1122--1135, Mar. 2016.
doi: 10.1016/j.cell.2016.02.038

\bibitem{Schwarz19}
A. Schwarz et al,
\emph{Systematic Review on Kinematic Assessments of Upper Limb Movements After Stroke},
\emph{Stroke},
vol. 50, no. 3, pp. 718--727, Mar. 2019.
doi: 10.1161/strokeaha.118.023531

\bibitem{Schwarz21}
A. Schwarz et al,
\emph{Measures of Interjoint Coordination Post-stroke Across Different Upper Limb Movement Tasks},
\emph{Frontiers in Bioengineering and Biotechnology},
vol. 8, Jan. 2021.
doi: 10.3389/fbioe.2020.620805

\bibitem{Singer17}
B. Singer and J. Garcia-Vega,
\emph{The Fugl-Meyer upper extremity scale},
\emph{Journal of physiotherapy},
vol. 63, no. 1, p. 53, 2017.

\bibitem{Ueyama21}
Y. Ueyama,
\emph{Costs of position, velocity, and force requirements in optimal control induce triphasic muscle activation during reaching movement},
\emph{Scientific Reports},
vol. 11, no. 1, Aug. 2021.
doi: 10.1038/s41598-021-96084-2

\bibitem{Valevicius19}
A. M. Valevicius et al,
\emph{Characterization of normative angular joint kinematics during two functional upper limb tasks},
\emph{Gait and Posture},
vol. 69, pp. 176--186, Mar. 2019.
doi: 10.1016/j.gaitpost.2019.01.037

\bibitem{Valevicius18}
A. M. Valevicius et al,
\emph{Use of optical motion capture for the analysis of normative upper body kinematics during functional upper limb tasks: A systematic review},
\emph{Journal of Electromyography and Kinesiology},
vol. 40, pp. 1--15, Jun. 2018.
doi: 10.1016/j.jelekin.2018.02.011

\bibitem{Valk19}
T. A. Valk et al,
\emph{Synergies reciprocally relate end-effector and joint-angles in rhythmic pointing movements},
\emph{Scientific Reports},
vol. 9, no. 1, Nov. 2019.
doi: 10.1038/s41598-019-53913-9

\bibitem{Yang18}
C. Yang et al,
\emph{Changes in movement variability and task performance during a fatiguing repetitive pointing task},
\emph{Journal of Biomechanics},
vol. 76, pp. 212--219, Jul. 2018.
doi: 10.1016/j.jbiomech.2018.05.025

\bibitem{McCrea05}
P. H. McCrea and J. J. Eng,
\emph{Consequences of increased neuromotor noise for reaching movements in persons with stroke},
\emph{Experimental Brain Research},
vol. 162, pp. 70--77, 2005.

\bibitem{Cirstea00}
M. C. Cirstea and M. F. Levin,
\emph{Compensatory strategies for reaching in stroke},
\emph{Brain},
vol. 123, no. 5, pp. 940--953, 2000.

\bibitem{Cacioppo20}
M. Cacioppo et al,
\emph{A new child-friendly 3D bimanual protocol to assess upper limb movement in children with unilateral cerebral palsy: Development and validation},
\emph{Journal of Electromyography and Kinesiology},
vol. 55, pp. 102481, Dec. 2020.
doi: 10.1016/j.jelekin.2020.102481

\bibitem{Corona18}
F. Corona et al,
\emph{Quantitative assessment of upper limb functional impairments in people with Parkinson's disease},
\emph{Clinical Biomechanics},
vol. 57, pp. 137--143, Aug. 2018.
doi: 10.1016/j.clinbiomech.2018.06.019

\bibitem{Woytowicz17}
E. J. Woytowicz et al,
\emph{Determining Levels of Upper Extremity Movement Impairment by Applying a Cluster Analysis to the Fugl-Meyer Assessment of the Upper Extremity in Chronic Stroke},
\emph{Archives of Physical Medicine and Rehabilitation},
vol. 98, no. 3, pp. 456--462, Mar. 2017.
doi: 10.1016/j.apmr.2016.06.023

\bibitem{Spilker97}
  J. Spilker, et al.,
  \textit{Using the NIH Stroke Scale to assess stroke patients},
  \textit{Journal of Neuroscience Nursing},
  vol. 29, no. 6, pp. 384--393, 1997.
  Publisher: Lippincott Williams \& Wilkins, WK Health.

\bibitem{Henarejos18}
  A. B. Meseguer-Henarejos, et al.,
  \textit{Inter- and intra-rater reliability of the Modified Ashworth Scale: a systematic review and meta-analysis},
  \textit{European Journal of Physical and Rehabilitation Medicine},
  vol. 54, no. 4, pp. 1--15, 2018.
  DOI: 10.23736/s1973-9087.17.04796-7.
  Publisher: Edizioni Minerva Medica.

\bibitem{Schwartz08}
  M. H. Schwartz, A. Rozumalski,
  \textit{The gait deviation index: A new comprehensive index of gait pathology},
  \textit{Gait and Posture},
  vol. 28, no. 3, pp. 351--357, 2008.
  DOI: 10.1016/j.gaitpost.2008.05.001.
  Publisher: Elsevier BV.

\bibitem{Baker09}
  R. Baker, et al.,
  \textit{The Gait Profile Score and Movement Analysis Profile},
  \textit{Gait and Posture},
  vol. 30, no. 3, pp. 265--269, 2009.
  DOI: 10.1016/j.gaitpost.2009.05.020.
  Publisher: Elsevier BV.

\bibitem{FMmethod}
  A. R. Fugl-Meyer, et al.,
  \textit{A method for evaluation of physical performance},
  \textit{Scand. J. Rehabil. Med},
  vol. 7, no. 1, pp. 13--31, 1975.
  Publisher: Scandinavian University Press.

\bibitem{Duncan94}
  P. W. Duncan, et al.,
  \textit{Similar motor recovery of upper and lower extremities after stroke},
  \textit{Stroke},
  vol. 25, no. 6, pp. 1181--1188, 1994.
  Publisher: American Heart Association.

\bibitem{Harrington20}
  R. M. Harrington, et al.,
  \textit{Roles of lesioned and nonlesioned hemispheres in reaching performance poststroke},
  \textit{Neurorehabilitation and neural repair},
  vol. 34, no. 1, pp. 61--71, 2020.
  Publisher: SAGE Publications.

\bibitem{Folstein75}
  M. F. Folstein, S. E. Folstein, and P. R. McHugh,
  \textit{"Mini-mental state": a practical method for grading the cognitive state of patients for the clinician},
  \textit{Journal of psychiatric research},
  vol. 12, no. 3, pp. 189--198, 1975.
  Publisher: Pergamon.

\bibitem{Sing13}
  G. C. Sing, et al.,
  \textit{Limb motion dictates how motor learning arises from arbitrary environmental dynamics},
  \textit{Journal of neurophysiology},
  vol. 109, no. 10, pp. 2466--2482, 2013.
  Publisher: American Physiological Society Bethesda, MD.

\bibitem{Bai14}
  L. Bai, et al.,
  \textit{Quantitative assessment of upper limb motion in neurorehabilitation utilizing inertial sensors},
  \textit{IEEE Transactions on Neural Systems and Rehabilitation Engineering},
  vol. 23, no. 2, pp. 232--243, 2014.
  Publisher: IEEE.

\bibitem{Ma17}
  H.-I. Ma et al.,
  \textit{Kinematic manifestation of arm-trunk performance during symmetric bilateral reaching after stroke: within vs. beyond arm’s length},
  \textit{American Journal of Physical Medicine \& Rehabilitation},
  vol. 96, no. 3, pp. 146--151, 2017.
  Publisher: LWW.

\bibitem{Mohapatra16}
  S. Mohapatra, R. Harrington, E. Chan, A. W. Dromerick, E. Y. Breceda, and M. Harris-Love,
  \textit{Role of contralesional hemisphere in paretic arm reaching in patients with severe arm paresis due to stroke: a preliminary report},
  \textit{Neuroscience letters},
  vol. 617, pp. 52--58, 2016.
  Publisher: Elsevier.

\bibitem{Wagner08}
  J. M. Wagner, J. A. Rhodes, and C. Patten,
  \textit{Reproducibility and minimal detectable change of three-dimensional kinematic analysis of reaching tasks in people with hemiparesis after stroke},
  \textit{Physical therapy},
  vol. 88, no. 5, pp. 652--663, 2008.
  Publisher: Oxford University Press.

\bibitem{Finley12}
  M. Finley et al.,
  \textit{Comparison of "less affected limb" reaching kinematics in individuals with chronic stroke and healthy age-matched controls},
  \textit{Physical \& Occupational Therapy in Geriatrics},
  vol. 30, no. 3, pp. 245--259, 2012.
  Publisher: Taylor \& Francis.

\bibitem{Patterson11}
  T. S. Patterson et al.,
  \textit{Reliability of upper extremity kinematics while performing different tasks in individuals with stroke},
  \textit{Journal of motor behavior},
  vol. 43, no. 2, pp. 121--130, 2011.
  Publisher: Taylor \& Francis.

\bibitem{Van14}
  L. van Dokkum et al.,
  \textit{The contribution of kinematics in the assessment of upper limb motor recovery early after stroke},
  \textit{Neurorehabilitation and neural repair},
  vol. 28, no. 1, pp. 4--12, 2014.
  Publisher: Sage Publications Sage CA: Los Angeles, CA.

\bibitem{Murphy11}
  M. A. Murphy et al.,
  \textit{Kinematic variables quantifying upper-extremity performance after stroke during reaching and drinking from a glass},
  \textit{Neurorehabilitation and neural repair},
  vol. 25, no. 1, pp. 71--80, 2011.
  Publisher: SAGE Publications Sage CA: Los Angeles, CA.

\bibitem{Sturm15}
  J.~Sturm, C.~Kerl, and D.~Cremers,
  \emph{Lecture (MOOC on EdX): Autonomous Navigation for Flying Robots (AUTONAVx)},
  2015,
  Technical University of Munich (TUM) / Metaio GmbH, Munich, Germany

\bibitem{NIST}
  W.~F.~Guthrie,
  \emph{NIST/SEMATECH e-Handbook of Statistical Methods (NIST Handbook 151)},
  National Institute of Standards and Technology,
  2020,
  Chapter 7.2.1.3.,
  Language: English,
  Copyright: License Information for NIST data

\bibitem{Wu05}
  G.~Wu et al,
  \emph{ISB recommendation on definitions of joint coordinate systems of various joints for the reporting of human joint motion—Part II: shoulder, elbow, wrist and hand},
  \emph{Journal of biomechanics},
  vol.~38,
  no.~5,
  pp.~981--992,
  year~2005,
  publisher: Elsevier.

\bibitem{Guzik22}
  A.~Guzik-Kopyto et al,
  \emph{Selection of Kinematic and Temporal Input Parameters to Define a Novel Upper Body Index Indicator for the Evaluation of Upper Limb Pathology},
  \emph{Applied Sciences},
  vol.~12,
  no.~22,
  p.~11634,
  year~2022,
  publisher: MDPI.

\bibitem{Hill22}
  S.~W.~Hill, S.~Mong, and Q.~Vo,
  \emph{Three-Dimensional Motion Analysis for Occupational Therapy Upper Extremity Assessment and Rehabilitation: A Scoping Review},
  \emph{The Open Journal of Occupational Therapy},
  vol.~10,
  no.~4,
  pp.~1--14,
  year~2022,
  publisher: Department of Occupational Therapy in the College of Health and Human~…

\bibitem{Wolff22}
  A.~Wolff, A.~Sama, M.~Lenhoff, and A.~Daluiski,
  \emph{The use of wearable inertial sensors effectively quantify arm asymmetry during gait in children with unilateral spastic cerebral palsy},
  \emph{Journal of Hand Therapy},
  vol.~35,
  no.~1,
  pp.~148--150,
  year~2022,
  publisher: Elsevier.

\bibitem{Cerveri07}
  P.Cerveri, et al.,
  \emph{Finger kinematic modeling and real-time hand motion estimation},
  \emph{Annals of biomedical engineering},
  vol.~35,
  pp.~1989--2002,
  year~2007,
  publisher: Springer.

\bibitem{Dehbandi16}
  B.Dehbandi, et al.,
  \emph{Using data from the Microsoft Kinect 2 to quantify upper limb behavior: a feasibility study},
  \emph{IEEE journal of biomedical and health informatics},
  vol.~21,
  no.~5,
  pp.~1386--1392,
  year~2016,
  publisher: IEEE.

\bibitem{Riad11}
J. Riad, S. Coleman, D. Lundh, and E. Brostr{\"o}m,
\textit{Arm posture score and arm movement during walking: a comprehensive assessment in spastic hemiplegic cerebral palsy},
\textit{Gait \& Posture}, vol. 33, no. 1, pp. 48--53, 2011.

\bibitem{saenen22}
L. Saenen, J.-J. Orban de Xivry, and G. Verheyden,
\textit{Development and Validation of a Novel Robot-Based Assessment of Upper Limb Sensory Processing in Chronic Stroke},
\textit{Brain Sciences}, vol. 12, no. 8, p. 1005, 2022.

\bibitem{Karamanidis16}
K. Karamanidis, A. Arampatzis, and G.-P. Br{\"u}ggemann,
\textit{Symmetry and reproducibility of kinematic parameters during various running techniques},
\textit{ISBS-Conference Proceedings Archive}, 2016.

\bibitem{rida19}
I. Rida, N. Almaadeed, and S. Almaadeed,
\textit{Robust gait recognition: a comprehensive survey},
\textit{IET Biometrics}, vol. 8, no. 1, pp. 14--28, 2019.

\bibitem{wong19}
A. L. Wong et al.,
\textit{Movement imitation via an abstract trajectory representation in dorsal premotor cortex},
\textit{Journal of Neuroscience}, vol. 39, no. 17, pp. 3320--3331, 2019.

\bibitem{passos17}
P. Passos, T. Campos, and A. Diniz,
\textit{Quantifying the degree of movement dissimilarity between two distinct action scenarios: An exploratory approach with Procrustes analysis},
\textit{Frontiers in Psychology}, vol. 8, p. 640, 2017.

\bibitem{Kamal18}
K. Sehairi, F. Chouireb, and J. Meunier,
\textit{Elderly fall detection system based on multiple shape features and motion analysis},
in \textit{2018 International Conference on Intelligent Systems and Computer Vision (ISCV)}, 2018, pp. 1--8. doi: 10.1109/ISACV.2018.8354084.

\bibitem{Anwary19}
A. R. Anwary, H. Yu, and M. Vassallo,
\textit{Gait Evaluation Using Procrustes and Euclidean Distance Matrix Analysis},
\textit{IEEE Journal of Biomedical and Health Informatics}, vol. 23, no. 5, pp. 2021-2029, 2019. doi: 10.1109/JBHI.2018.2875812.

\bibitem{seber09}
G. A. F. Seber,
\textit{Multivariate Observations},
John Wiley \& Sons, 2009.

\bibitem{bookstein97}
F. L. Bookstein,
\textit{Morphometric Tools for Landmark Data},
Cambridge University Press, Cambridge, UK, 1997.

\bibitem{kendall89}
D. G. Kendall,
\textit{A Survey of the Statistical Theory of Shape},
\textit{Statistical Science}, vol. 4, no. 2, pp. 87-99, 1989, Institute of Mathematical Statistics.

\bibitem{Aron19}
A. Aron, E. N. Aron, and E. J. Coups,
\textit{Statistics for Psychology},
7th edition, Pearson, 2019.

\bibitem{Wolff23}
A. L. Wolff et al,
\textit{Dynamic Assessment of the Upper Extremity: A Review of Available and Emerging Technologies},
\textit{Journal of Hand Surgery (European Volume)}, 2023, SAGE Publications Sage UK: London, England.

\bibitem{Passos23}
P. Passos et al,
\textit{Improvements (or not!) in the Hand Trajectory of Stroke Patients Due to Practice of a Virtual Game},
\textit{Adaptive Behavior}, pp. 10597123231163595, 2023, SAGE Publications Sage UK: London, England.

\end{thebibliography}
\end{document}